\documentclass[12pt,a4paper]{article}
\usepackage{epsfig}
\begin{document}
\textwidth=135mm
 \textheight=200mm
\begin{center}
{\bfseries Three-dimensional kaon and pion emission source extraction from $\sqrt {s_{NN}}=200$~GeV Au+Au collisions at RHIC-STAR 
\footnote{{\small Talk at the workshop "Workshop on Particle Correlations and Fluctuations 2010", BITP, Kiev, September 14 - 18,
2010.}}}
\vskip 5mm
P. Chung$^{\dag}$ for The STAR Collaboration
\vskip 5mm
{\small {\it $^\dag$ Nuclear Physics Institute ASCR, Prague, Czech Republic}} \\

\end{center}
\vskip 5mm
\centerline{\bf Abstract}

Three-dimensional source images for mid-rapidity, low transverse momentum kaon and pion pairs have been extracted from central Au+Au collisions data at $\sqrt {s_{NN}}=200$~GeV by the STAR experiment at RHIC. The pion source function displays significant non-Gaussian features implying a finite pion emission duration. On the other hand, the kaon source function is essentially Gaussian, consistent with
instantaneous emission from the fireball.
\vskip 10mm
\section{\label{sec:intro}Introduction}
Two-particle interferometry is widely believed to be the most sensitive technique available to probe the space-time characteristics of the hot expanding fireball created in heavy ion experiments \cite{lis05}. The first extraction of the detailed shape of the 3D source function for pion pairs by the PHENIX experiment has permitted the decoupling of the spatial and temporal aspects of the expanding fireball as well as the determination of a non-vanishing pion emission duration  using detailed model comparisons \cite{chu08}. However, a consistent picture of the underlying dynamics driving the expansion can be achieved only via systematic consideration of a variety of observables. Hence, this paper presents the space-time extent of the fireball as obtained from kaon interferometry and source imaging from STAR data under similar conditions as the pion source extracted by the PHENIX Collaboration.  

\section{Experimental Setup and Data Analysis}
  
The data presented here were taken by the STAR Collaboration during the year-2004 and 2007 runs. The colliding beams ($\sqrt {s_{NN}}= 200$ GeV) were provided by the RHIC accelerator.
Charged tracks were detected in the STAR Time Projection Chamber (TPC)
\cite{stardet}, covering $\pm$1.1 units of pseudo-rapidity, surrounded by a solenoidal magnet providing a nearly uniform magnetic field of 0.5 Tesla along the beam direction. The TPC provides tracking information for track reconstruction and particle identification by means of energy loss in the TPC gas.
 
The 3D correlation function, C($\mathbf{q}$), was calculated as the ratio of foreground
to background distributions in relative momentum $\mathbf{q}$ for $\pi^+\pi^+$ and $\pi^-\pi^-$ pairs as well as  $K^+K^+$ and $K^-K^-$ pairs. Here, $\mathbf{q}=\frac{(\mathbf{p_1}-\mathbf{p_2})}{2}$ is half of the relative
momentum between the two particles in the pair C.M. system (PCMS) frame. The foreground
distribution was obtained using pairs of particles from the same event and the background
was obtained by pairing particles from different events. Track
merging and splitting effects were removed by appropriate cuts in the
relevant coordinate space on both the foreground and background distributions \cite{adam05}.
There was no significant effect on the correlation function due to the
momentum resolution of 1$\%$ for the low transverse momentum under consideration.

In the Cartesian surface-spherical harmonic decomposition technique \cite{dan-chu05}, the 3D correlation function is expressed as
\begin{equation}
C(\mathbf{q}) - 1 = R(\mathbf{q}) = \sum_l \sum_{\alpha_1 \ldots \alpha_l}
   R^l_{\alpha_1 \ldots \alpha_l}(q) \,A^l_{\alpha_1 \ldots \alpha_l} (\Omega_\mathbf{q})
 \label{eqn1}
\end{equation}
where $l=0,1,2,\ldots$, $\alpha_i=x, y \mbox{ or } z$, $A^l_{\alpha_1 \ldots \alpha_l}(\Omega_\mathbf{q})$
are Cartesian harmonic basis elements ($\Omega_\mathbf{q}$ is solid angle in $\mathbf{q}$ space) and $R^l_{\alpha_1 \ldots \alpha_l}(q)$ are Cartesian correlation moments given by
\begin{equation}
 R^l_{\alpha_1 \ldots \alpha_l}(q) = \frac{(2l+1)!!}{l!}
 \int \frac{d \Omega_\mathbf{q}}{4\pi} A^l_{\alpha_1 \ldots \alpha_l} (\Omega_\mathbf{q}) \, R(\mathbf{q})
 \label{eqn2}
\end{equation}
The coordinate axes are oriented so that $z$ is parallel to the beam (long) direction, $x$ points
in the direction of the total momentum of the pair in the Locally Co-Moving System (out) and $y$ (side) is perpendicular to $x$ and $z$.
  
The correlation moments, for each order $l$, can be calculated from the measured 3D correlation function using Eq.~(\ref{eqn2}). In this analysis, Eq.~(\ref{eqn1}) is truncated at $l=6$ for pions and at $l=4$ for kaons and expressed in terms of independent moments only. Higher order moments were found
to be negligible. Up to order 6, there are 10 independent moments:
$R^0$, $R^2_{x2}$, $R^2_{y2}$, $R^4_{x4}$, $R^4_{y4}$, $R^4_{x2y2}$,
$R^6_{x6}$, $R^6_{y6}$, $R^6_{x4y2}$ and $R^6_{x2y4}$
where $R^2_{x2}$ is shorthand for $R^2_{xx}$ etc. These independent moments were extracted as a function of $q$, by fitting the truncated series to
the measured 3D correlation function with the moments as the parameters of
the fit.

 Each independent correlation moment is then imaged using the 1D Source Imaging code of Brown and Danielewicz \cite{brown97-98} to obtain the corresponding source moment for each order $l$. Bose-Einstein symmetrisation and Coulomb interaction (the sources of the observed correlations) are contained in the source imaging code.
Thereafter, the total source function is constructed by combining the source
moments for each $l$ as in Eq.~(\ref{eqn3})
  
\begin{equation}
 S(\mathbf{r}) = \sum_l \sum_{\alpha_1 \ldots \alpha_l}
   S^l_{\alpha_1 \ldots \alpha_l}(r) \,A^l_{\alpha_1 \ldots \alpha_l} (\Omega_\mathbf{r})
\label{eqn3}
\end{equation}

Alternatively, the 3D source function can also be extracted by directly fitting 
the 3D correlation function with an assumed functional form for $S(\mathbf r)$. This corresponds to a simultaneous fit of the independent moments with the assumed functional form via the 3D Koonin-Pratt equation Eq.~(\ref{3dkpeqn}) 

\begin{equation}
  C(\mathbf{q})-1 = R(\mathbf{q}) = \int d\mathbf{r} K(\mathbf{q},\mathbf{r}) S(
\mathbf{r})
  \label{3dkpeqn}
\end{equation}

\section{Results}

\begin{figure}
\includegraphics[width=0.5\linewidth]{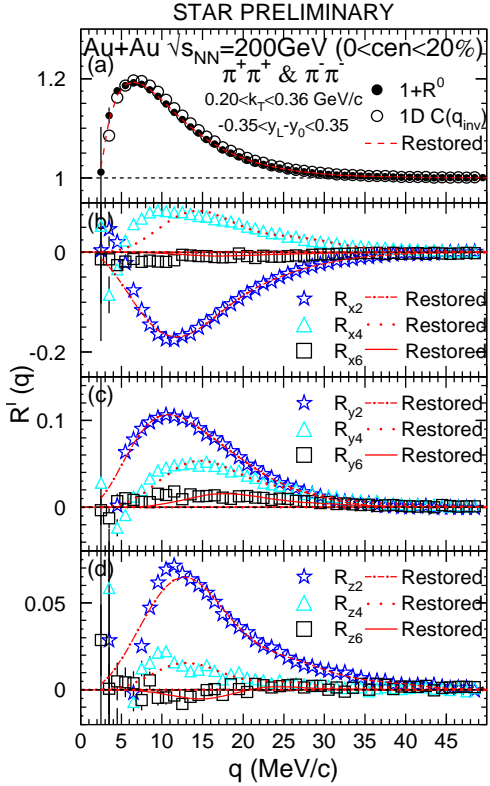}
\vskip -0.5cm
\caption{\small{Pion correlation moments (open symbols) for multipolarity orders (a) l=0 (b) l=2 (c) l=4 and (d) l=6. Restored moments are shown as curves. Panel (a) also shows a comparison between $R^0(q)$ and $R(q)=C(q)-1$.}}
\label{pion_moments_star}
  
\end{figure}

Figure~\ref{pion_moments_star} (a) compares the $l=0$ moment (solid circle) with the 1D correlation function (open circle), for mid-rapidity ($|y|<0.35$, where $y$ is particle rapidity), low tranverse momentum $k_T$
($0.2<k_T<0.36$~GeV/c) $\pi^+\pi^+$ and $\pi^-\pi^-$pairs from 200~AGeV central (20$\%$) Au+Au collisions, as a function of the relative momentum $q$ in the PCMS. They are in very close agreement, as is expected in the absence of significant angular acceptance issues.

The model-independent imaging technique of Brown and Danielewicz 
\cite{brown97-98} is used to image each relevant correlation moment to obtain the corresponding source moment. The latter are then combined according to Eq.~\ref{eqn3} to yield the 3D source function. 

Figure~\ref{pion_moments_star} (b), (c) and (d) compares the magnitude of the
correlation moments (symbols), for increasing orders $l$, in the $x$, $y$ and $z$ directions respectively. The moments decrease in magnitude with
increasing order and are virtually 0 for $l=6$. This justifies truncating
the series Eq.~(\ref{eqn1}) at $l=6$.

The restored correlation moments, obtained from imaging the actual correlation moments, are shown by lines in Figure~\ref{pion_moments_star} (b)-(d) for $l=0, 2, 4, 6$. They are in good agreement with the actual correlation moments, implying that the imaging process was well under control.

\begin{figure}
\includegraphics[width=0.5\linewidth]{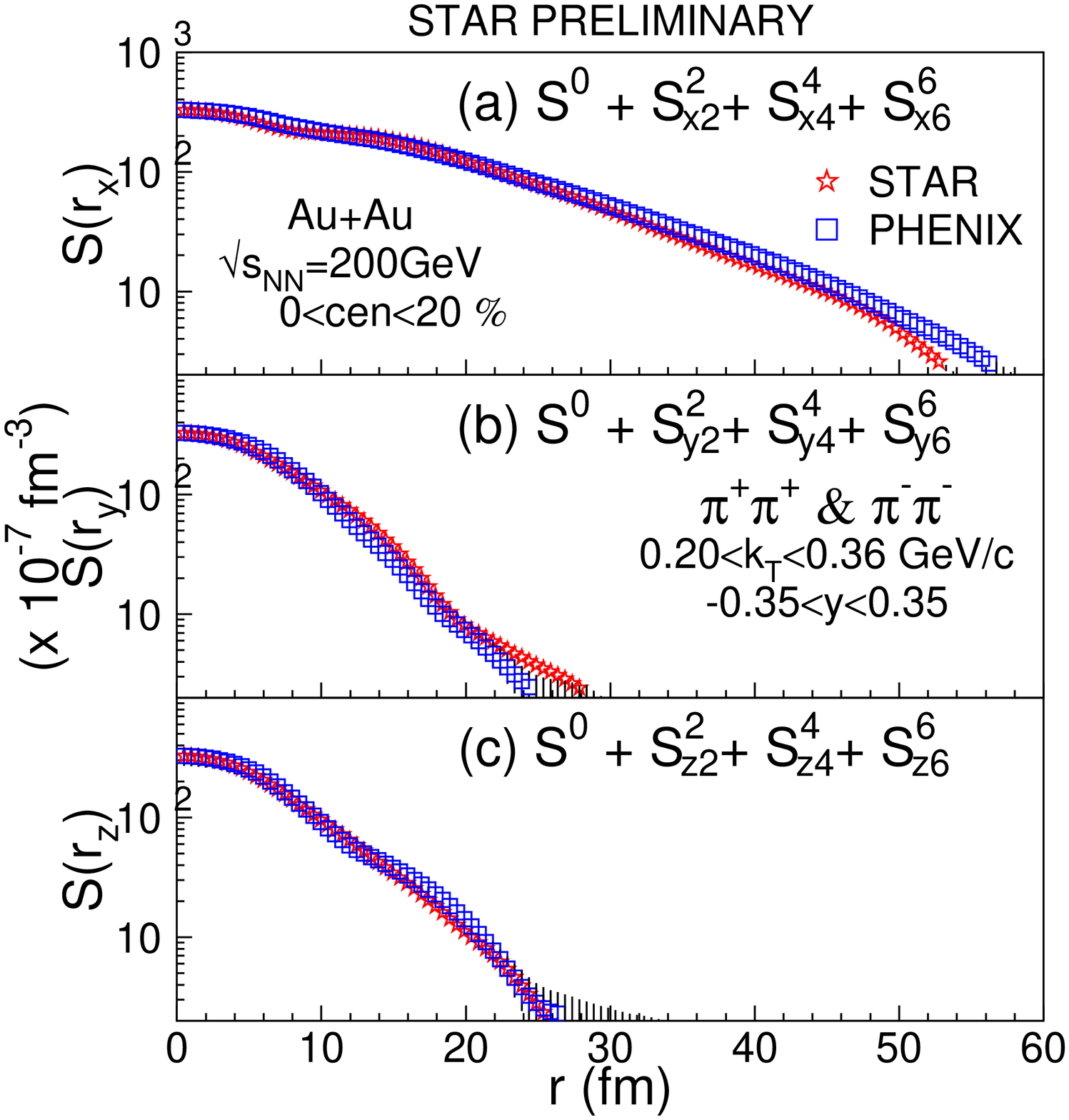}
\includegraphics[width=0.5\linewidth]{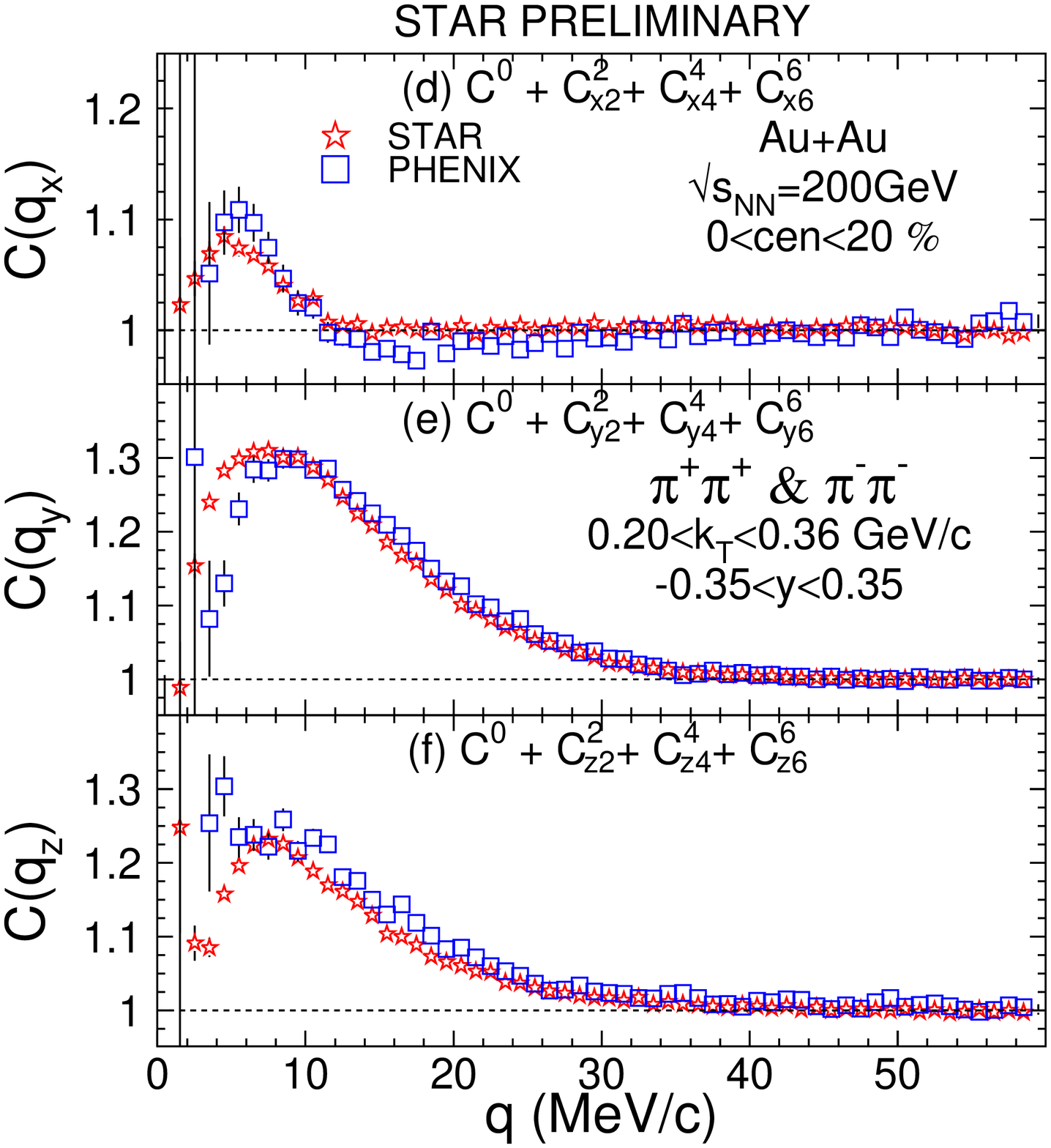}
\vskip -1.5cm
\caption{\small{Comparison of pion source function profiles S$(r_x)$, S$(r_y)$ and S$(r_z)$ (panels (a)-(c)) and their associated correlation profiles C$(q_x)$, C$(q_y)$ and C$(q_z)$ (panels (d)-(f)) in the PCMS from the STAR (stars) and PHENIX (squares) experiments.}}
\label{pion_src_cor_star_phnx}
\end{figure}

Figure~\ref{pion_src_cor_star_phnx} (a)-(c) show profiles of the extracted 3D source function in the $x$, $y$ and $z$ directions obtained by summing the source moments
up to $l=6$. The source image (stars) is in good agreement with that reported by the PHENIX Collaboration (squares) in all 3 directions \cite{chu08}. Panels (d)-(f) show that the corresponding correlation function profiles from STAR and PHENIX are also in good agreement. One can also note that the broader source profile in the $x$ direction is associated with the narrower correlation profile. 

\begin{figure}
\includegraphics[width=0.495\linewidth]{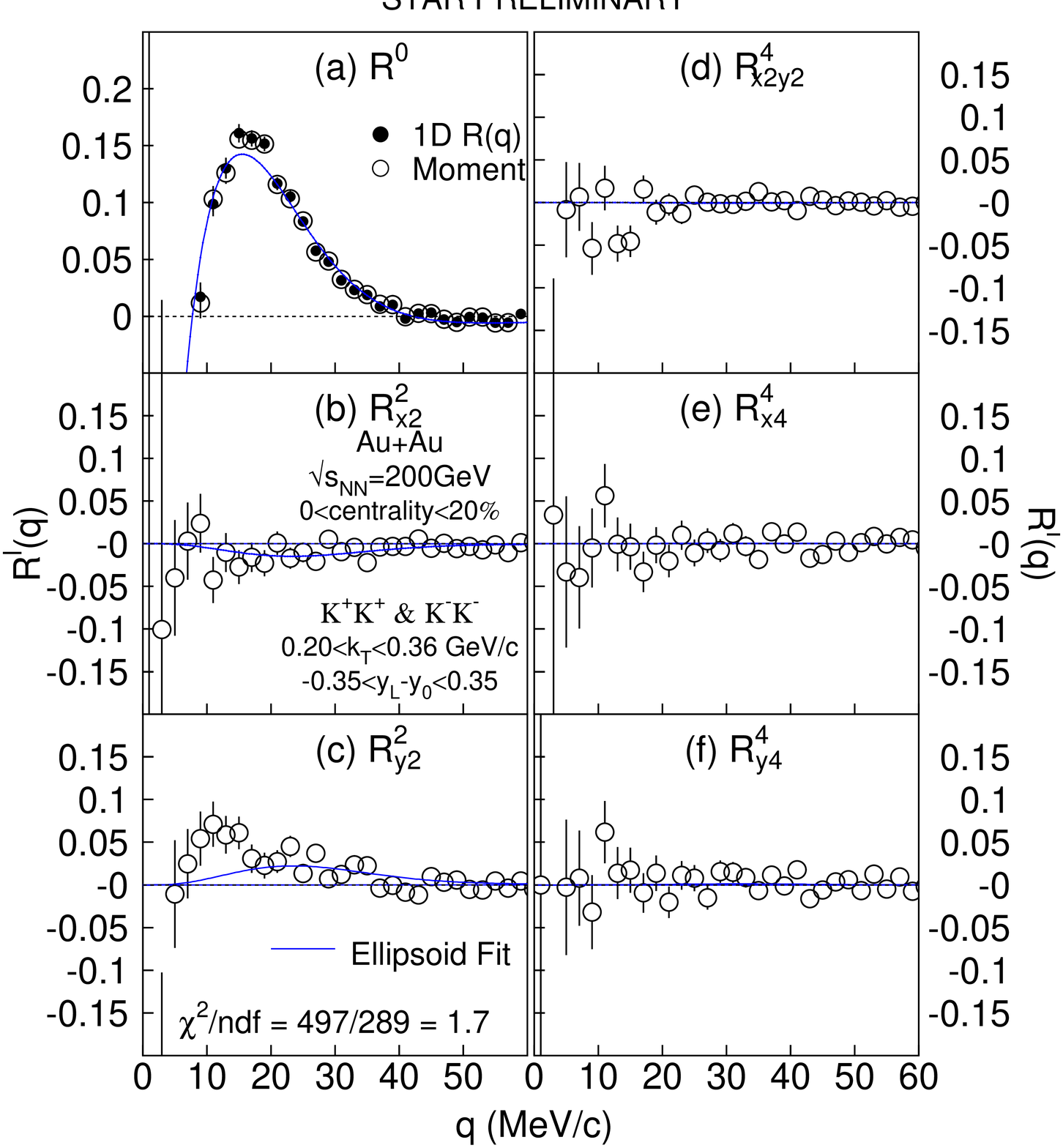}
\includegraphics[width=0.495\linewidth]{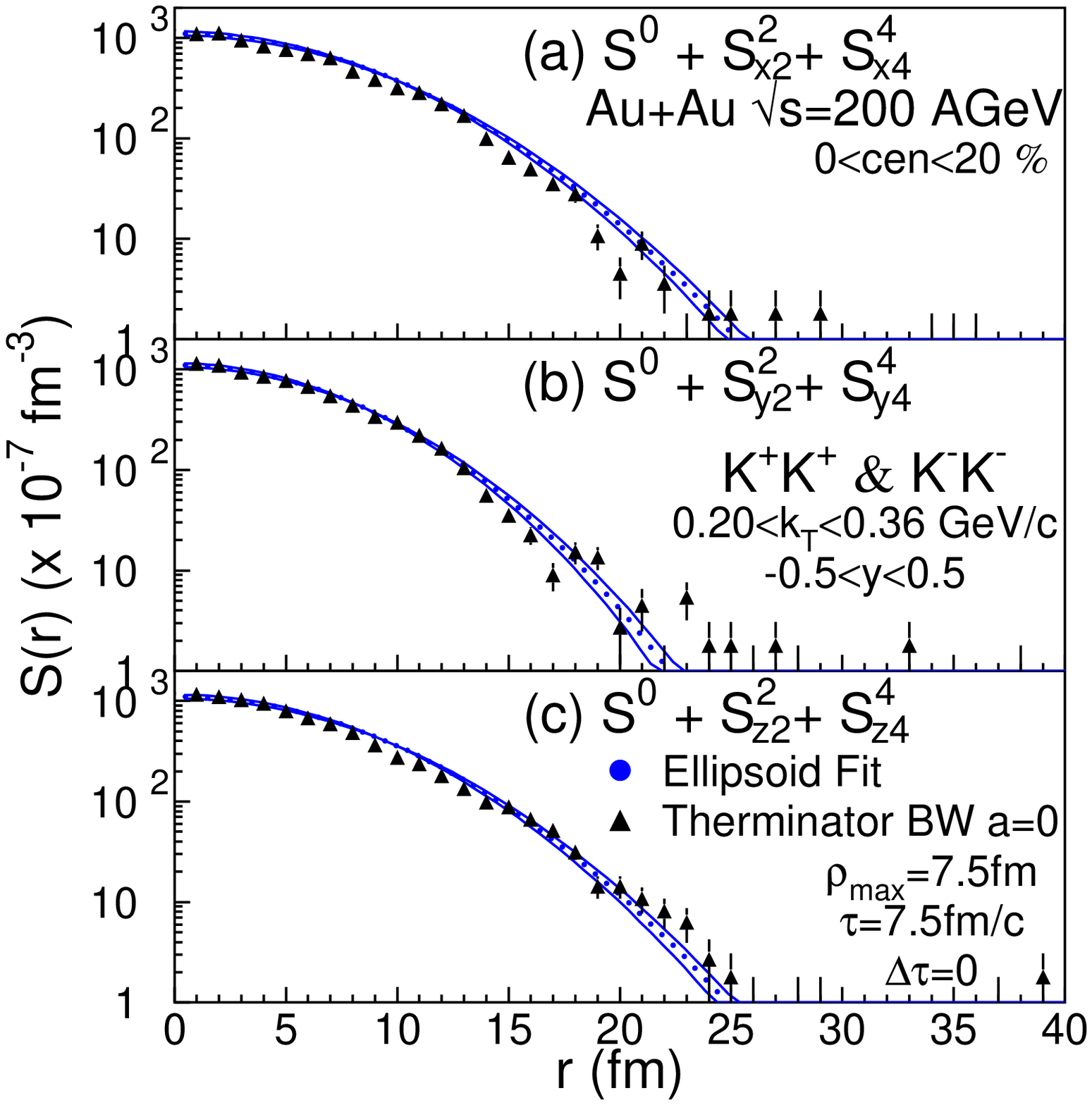}
\vskip -0.5cm
\caption{\small{Panels (a)-(f) left: Correlation moments $R^l(q)$ for $l=$0, 2, 4. Panel (a) also shows a comparison between $R^0(q)$ and 1D $R(q)$. The solid lines indicate the ellipsoid fit. Panels (a)-(c) right: Kaon source function profiles from STAR. Solid triangles depict Therminator model calculation with indicated parameter values.}}
\label{kaon_moments_star}
\end{figure}

Figure~\ref{kaon_moments_star} (a)-(f) (left) shows the correlation moments (open circles) obtained from the cartesian surface-spherical harmonic decomposition of the 3D correlation function for mid-rapidity ($|y|<0.5$), low $k_T$ ($0.2<k_T<0.36$~GeV/c) $K^+K^+$ and $K^-K^-$pairs from 200~AGeV central (20$\%$) collisions. Panel (a) shows that the $l=0$ moment (open circle) is in good agreement with the 1D correlation function (solid circle). The $l=2$ moments in panels (b)-(c) are relatively small and the $l=4$ moments in panels (d)-(f) are statistically consistent with 0. The curve represents an ellipsoid fit to the six independent moments. The $\chi^2/ndf$ value of 1.7 indicates that the kaon source function is reasonably described by an ellipsoidal shape.  

The dotted curve in figure~\ref{kaon_moments_star} (a)-(c) (right) shows the profile of the ellipsoid fit in the x, y and z directions. The band around the curve indicates the statistical uncertainty on the source fit. The extracted kaon source function is fairly well reproduced by the Therminator model \cite{kis05-07} calculation in blast-wave mode (triangles) with a transverse size of $\rho_{max}=7.5$~fm and lifetime $\tau=7.5$~fm/c of the emitting fireball and kaon emission duration $\Delta\tau=0$.

\section{Discussions}

The difference in the extracted 3D source function between pions and kaons coupled with the determination of similar spatio-temporal characteristics of the expanding fireball from both pion and kaon correlation functions suggests that pion femtoscopy is significantly influenced by resonance decays and emission duration effects compared to that for kaon. Indeed, a non-vanishing pion emission duration of about 2 fm/c was reported by the PHENIX Collaboration \cite{chu08} whereas the new STAR result points to kaon emission as being essentially instantaneous within the framework of the Therminator model . Furthermore, the kinetic freeze-out process for pions is different from that for kaons as evidenced by the different values for the slope parameter, $a$, of the freeze-out hypersurface. The latter specifies how the freeze-out time varies with the transverse radius of the expanding fireball. While a negative value for $a$ was necessary for pion emission, implying emission of pions from outside-in, a zero value for $a$ is sufficient here to explain kaon emission.

\section{Acknowledgments}

This work was supported by grant LC07048 of MSMT of the Czech Republic.

\end{document}